\documentclass[a4paper,clock]{article}
\usepackage[dvips]{epsfig}
\usepackage{amssymb}
\usepackage{amsmath}
\usepackage{bm}
\usepackage{dcolumn}

\catcode`\@=11
\def\marginnote#1{}
\newcount\hour
\newcount\minute
\newtoks\amorpm
\hour=\time\divide\hour by60
\minute=\time{\multiply\hour by60 \global\advance\minute
by-\hour}\edef\standardtime{{\ifnum\hour<12
\global\amorpm={am}%
        \else\global\amorpm={pm}\advance\hour by-12 \fi
        \ifnum\hour=0 \hour=12 \fi
        \number\hour:\ifnum\minute<10
0\fi\number\minute\the\amorpm}}
\edef\militarytime{\number\hour:\ifnum\minute<10
0\fi\number\minute}

\def\draftlabel#1{{\@bsphack\if@filesw {\let\thepage\relax
   \xdef\@gtempa{\write\@auxout{\string
      \newlabel{#1}{{\@currentlabel}{\thepage}}}}}\@gtempa
   \if@nobreak \ifvmode\nobreak\fi\fi\fi\@esphack}
        \gdef\@eqnlabel{#1}}
\def\@eqnlabel{}
\def\@vacuum{}
\def\draftmarginnote#1{\marginpar{\raggedright\scriptsize\tt#1}}
\def\draft{\oddsidemargin -.5truein
        \def\@oddfoot{\sl preliminary draft \hfil
        \rm\thepage\hfil\sl\today\quad\militarytime}
        \let\@evenfoot\@oddfoot \overfullrule 3pt
        \let\label=\draftlabel
        \let\marginnote=\draftmarginnote

\def\@eqnnum{(\theequation)\rlap{\kern\marginparsep\tt\@eqnlabel}%
\global\let\@eqnlabel\@vacuum}  }

\def\numberbysection{\@addtoreset{equation}{section}
        \def\theequation{\thesection.\arabic{equation}}}

\def\underline#1{\relax\ifmmode\@@underline#1\else
 $\@@underline{\hbox{#1}}$\relax\fi}

\catcode`@=12
\relax

\numberbysection

\advance \topmargin by -\headheight
\advance \topmargin by -\headsep

\evensidemargin \oddsidemargin
\def\br{\begin{eqnarray}}
\def\er{\end{eqnarray}}
\def\be{\begin{equation}}
\def\ee{\end{equation}}

\def\({\left(}
\def\){\right)}

\relax

\def\a{\alpha}

\def\b{\beta}

\def\d{\delta}
\def\D{\Delta}

\def\g{\gamma}
\def\G{\Gamma}

\def\l{\lambda}

\def\O{\Omega}

\def\pa{\partial}

\def\tp0{\Theta_{+}^{(0)}}
\def\tm0{\Theta_{-}^{(0)}}

\def\f#1#2#3 {f^{#1#2}_{#3}}

\def\win1{{\sf w_{1+\infty}}}

\def\Win1{{\sf W_{1+\infty}}}

\def\rlx{\relax\leavevmode}
\def\inbar{\vrule height1.5ex width.4pt depth0pt}
\def\IZ{\rlx\hbox{\sf Z\kern-.4em Z}}
\def\IR{\rlx\hbox{\rm I\kern-.18em R}}
\def\IT{\rlx\hbox{\rm I\kern-.18em T}}
\def\IC{\rlx\hbox{\,$\inbar\kern-.3em{\rm C}$}}
\def\IN{\rlx\hbox{\rm I\kern-.18em N}}
\def\IO{\rlx\hbox{\,$\inbar\kern-.3em{\rm O}$}}
\def\IP{\rlx\hbox{\rm I\kern-.18em P}}
\def\IQ{\rlx\hbox{\,$\inbar\kern-.3em{\rm Q}$}}
\def\IF{\rlx\hbox{\rm I\kern-.18em F}}
\def\IG{\rlx\hbox{\,$\inbar\kern-.3em{\rm G}$}}
\def\IH{\rlx\hbox{\rm I\kern-.18em H}}
\def\II{\rlx\hbox{\rm I\kern-.18em I}}
\def\IK{\rlx\hbox{\rm I\kern-.18em K}}
\def\IL{\rlx\hbox{\rm I\kern-.18em L}}
\def\one{\hbox{{1}\kern-.25em\hbox{l}}}
\def\0#1{\relax\ifmmode\mathaccent"7017{#1}%
B        \else\accent23#1\relax\fi}

\def\PRL#1#2#3{{\sl Phys. Rev. Lett.} {\bf#1} (#2) #3}
\def\NPB#1#2#3{{\sl Nucl. Phys.} {\bf B#1} (#2) #3}

\def\CMP#1#2#3{{\sl Commun. Math. Phys.} {\bf #1} (#2) #3}

\def\PLA#1#2#3{{\sl Phys. Lett.} {\bf #1A} (#2) #3}

\def\PLB#1#2#3{{\sl Phys. Lett.} {\bf #1B} (#2) #3}

\def\JPA#1#2#3{{\sl J. Physics} {\bf A#1} (#2) #3}

\def\JHEP#1#2#3{{\sl JHEP} {\bf #1} (#2) #3}

\def\Nonl#1#2#3{{\sl Nonlinearity} {\bf #1} (#2) #3}

\def\CNSNS#1#2#3{{\sl Commun Nonlinear Sci Numer Simulat} {\bf #1} (#2) #3}
\def\ScR#1#2#3{{\sl Sci. Rep.} {\bf #1} (#2) #3}

\def\OQE#1#2#3{{\sl Optical and Quantum Electronics} {\bf #1} (#2) #3}
\def\IJMPB#1#2#3{{\sl  International Journal of Modern Physics B} {\bf #1} (#2) #3}
\begin{document}

\begin{titlepage}

\vspace{.2in}
\begin{center}
{\large\bf Asymptotically conserved charges and 2-kink collision in quasi-integrable potential KdV models}
\end{center}

\vspace{.2in}

\begin{center}

Harold Blas   

\par \vskip .2in \noindent

Instituto de F\'{\i}sica\\
Universidade Federal de Mato Grosso\\
Av. Fernando Correa, $N^{0}$ \, 2367\\
Bairro Boa Esperan\c ca, Cep 78060-900, Cuiab\'a - MT - Brazil 

\normalsize
\end{center}

\vspace{.3in}

\begin{abstract}
\vspace{.3in}
We study a particular deformation of the potential KdV model (pKdV) and construct the quasi-conservation laws by a direct method. The charge densities, differing from their integrable counterpart with homogeneous degree terms, exhibit mixed scale dimension terms. The modifications of the charges around the soliton interaction regions are examined by numerically simulating some representative anomalies. We show numerically the elastic scattering of two kinks for a wide range of values of the deformation parameters.  It is discussed an anomaly cancellation mechanism to define an exact conservation law of the usual pKdV model, and a renormalization procedure is introduced for some divergent charges by subtructing the continuous linear background contribution. The KdV-type equations are quite ubiquitous in several areas of non-linear science, such as the study of General Relativity in $AdS_{3}$, Bose-Einstein condensates, superconductivity and fluid dynamics. 
\end{abstract} 

\end{titlepage}

\section{Introduction}

Integrable models are characterized by soliton solutions and an infinite number of conserved charges, making them mathematically elegant and physically insightful \cite{das, faddeev, babelon}. However, many important physical systems with solitary wave solutions do not fall under the integrable framework. In this context, it has been put forward the quasi-integrability concept through the anomalous Lax equation \cite{jhep1, jhep2} and pseudo-potential approaches \cite{npb20, cnsns, jhep6},  to tackle the properties of modifications of integrable models.

The developments follow by the construction of an infinite number of asymptotically conserved charges and the examination of the space-time inversion symmetry property of the anomalies. These charges are asymptotically conserved as long as the space-time integral of the corresponding anomaly densities vanishes. The vanishing of the anomalies, in general, have been verified numerically, except for some sub-models with particular set of deformation parameters. Moreover, in the anomalous Lax  and pseudo-potential approaches the charge densities maintain a similar form to those of the corresponding undeformed theories, i.e. they exhibit terms with homogeneous scale dimensions as the integrable counterparts. 

Recent research has uncovered several new towers of infinite number of asymptotically conserved charges within deformed sine-Gordon, NLS and KdV models \cite{npb20, ijmpb1, ijmpb2, jhep6}. These new charges differ in form from those of the undeformed model, as they exhibit terms with mixed scale dimensions.  It is noted that a subset of these new charges are anomalous even for the standard integrable counterparts.

In this work we consider a deformation of the potential KdV model (pKdV) and  examine the type of charges, local and non-local,  which  depend explicitly on the deformation parameters and exhibit mixed scale dimension terms. The pKdV is one of the non-relativistic scalar field models which exhibit kink-type topological solitons supported by non-vanishing boundary conditions (nvbc). Regarding topological solitons it has been considered the deformations of the relativistic SG  and Bullough-Dodd models with kink solitons in \cite{jhep1} and \cite{auri}, respectively. So, it is interesting to consider the quasi-integrability properties of a scalar field supporting non-relativistic topological solitons, since such theories also appear in many areas of non-linear science, condensed matter physics, plasma physics, Bose-Einstein condensates and,
in particular, fluid dynamics. In this context, our aim is to
predict the results of solitary wave collisions and test the quasi-integrability concept in
deformed pKdV models with nvbc. However, the nvbc inherent to the kink solitons, which may change when the deformations are present, introduces new features when applying the techniques developed for deformed KdV bright solitons. For example, some quasi-conserved charges would require a renormalization procedure by subtructing the contribution of the continuous linear background, since the solution incorporates this vacuum solution plus the kink soliton itself.  

The space-time reflection symmetric ${\cal P}$ analytical N-kink solutions will be obtained for the undeformed pKdV model, and the  ${\cal P}$ symmetric 2-kink analytical solution for a sub-model, the so-called non-integrable potential modified regularized long wave model (pmRLW). Using the vanishing integrated anomaly for the 2-kink solutions we will show analytically the quasi-integrability of the pmRLW model.    

We perform the construction of the quasi-conservation laws by a direct approach starting from the equation of motion. The numerical simulation of two-soliton interactions in the deformed model sheds light on the dynamics of the system. The observed elastic collisions, where solitons maintain their shapes and velocities without significant radiation loss, hint at the robustness and coherence of soliton behavior even in non-integrable systems. Overall, this research contributes to our understanding of soliton dynamics in non-integrable systems with non-relativistic kink solitons; so, highlighting the persistence and interesting phenomena that emerge even in the absence of strict integrability.

The paper is organized as follows. The next section presents a particular deformation of the pKdV model. In sec. \ref{sec:quasi1} some local and non-local quasi-conservation laws are constructed through a direct approach. In sec. \ref{sec:tau1}, using the tau function method, we provide the pKdV analytical N-kink solitons, as well as the 1-kink soliton for the deformed pKdV. A direct method allows as to find a general 1-kink solution of the deformed pKdV. The 1-kink and 2-kink solitons of the  pmRLW theory are also uncovered. In sec. \ref{sec:dpkdv11} it is discussed the analytical quasi-integrability of the  pmRLW theory for the 2-kink solutions. In sec. \ref{sec:num} we numerically  simulate the vanishing of some representative anomalies of the deformed pKdV model, and in sec. \ref{sec:discuss} we discuss the results  and present the conclusions.
 
\section{A particular deformation of the pKdV model}
\label{sec:dpkdv}
 
We will consider a deformation of the pKdV model. It  involves the real scalar field $v$ and the auxiliary field $w$ with equation of motion
\br
\label{dpkdv}
v_t + v_x +\frac{\a}{2} v_x^2 + \, v_{xxx} &=& X,\\
X & \equiv & -\epsilon_2 \frac{\a}{4} w_x v_t  + \epsilon_1 (v_{xxt} + v_{xxx} ), \label{dpkdv1}
\er
such that the auxiliary field satisfies
\br
\label{wv1}
v_x &=& w_ t.
\er
The real numbers $\epsilon_1$  and  $\epsilon_2$ are the deformation parameters away from the usual pKdV and $\a$ is an arbitrary real parameter defining the nonlinear term of the model. The model (\ref{dpkdv}) embraces a variety of sub-models; e.g.  for $\epsilon_1=  \epsilon_2=0$ one has the integrable pKdV model, and for $\epsilon_1=\epsilon_2=1$ the non-integrable potential modified regularized long wave model (pmRLW). 

Note that defining $u \equiv \pa_x v$ one can write the $x-$derivative of the eq. (\ref{dpkdv}) as
\br
\label{dkdv}
u_t + u_x +\a u u_x + \, u_{xxx} &=& \pa_x X,\\ 
\nonumber
u &=& v_x = w_t.
\er
This is the so-called modified KdV model which has been studied in \cite{npb, jhep6} in the context of quasi-integrability. However, the model (\ref{dpkdv}) deserves to be studied separately in order to examine the quasi-integrability properties and  its kink-type solitons. In fact, the undeformed pKdV model ($X=0$) exhibits a kink-type soliton corresponding to a non-relativistic scalar model. So, it is  interesting to uncover its quasi-integrability properties, such as the quasi-conservation laws and the kink collisions for deformed models ($X \neq 0$).

A suitable parametrization of the model (\ref{dpkdv}) is available in order to construct analytical or numerical soliton solutions  of the model. So, let us consider
\br
\label{wxvt}
 w= -\frac{8}{\a} q_{x} \,\,\, \mbox{and}  \,\,\,v= -\frac{8}{\a} q_{t}.
\er 
So, substituting the expressions of $w$ and $v$  from (\ref{wxvt}), respectively,  into (\ref{dpkdv}) one gets 
\br
\label{eqq}
q_{tt} + q_{xt} - 4 q^2_{xt} - 2 \epsilon_2  q_{xx} q_{tt} + q_{xxxt} - \epsilon_1 (q_{xxtt} + q_{xxxt} ) = 0.
\er
Notice that this equation exhibits as the vacuum solution a continuous linear background of the form
\br
\label{clb}
q_{clb} = K x - \O\, t + c_o.
\er
So, a general 1-soliton solution will incorporate this vacuum solution plus the `kink' soliton itself. In fact,  the kink-type soliton for the field $v$  ($v \sim `kink' + \pa_t q_{clb} $) will interpolate two vacua of the form  (\ref{clb}).  

Next, let us discuss some space-time symmetries related to soliton-type solutions of the model. So, consider the space-time reflection around a given fixed point $(x_{\Delta},t_{\Delta})$
\br
\label{parity1}
{\cal P}:  (\widetilde{x},\widetilde{t}) \rightarrow (-\widetilde{x},-\widetilde{t});\,\,\,\,\,\,\,\,\widetilde{x} = x - x_{\Delta},\,\,\widetilde{t} = t- t_{\Delta}. 
\er 
In fact, the transformation ${\cal P}$ defines a shifted parity ${\cal P}_{s}$ for the spatial variable  and the  delayed time reversal ${\cal T}_d$ for the time variable. When $x_{\Delta}=0$ ($t_\Delta=0$), ${\cal P}_{s}$ (${\cal T}_d$) is reduced back to the pure parity ${\cal P}$ (pure time reversal ${\cal T}$).  
  
Following the quasi-integrability approach \cite{jhep6} let us assume that the $v-$field solution of the deformed pKdV model evaluated on the N-soliton solution, viz. $v_{N_{-}sol}$, under (\ref{parity1}) transforms as
\br
\label{paritys1}
{\cal P} (v_{N-sol})= - v_{N-sol} + const. 
\er 
This implies, according to (\ref{wv1}) and (\ref{wxvt}), that 
\br
\label{vwtr}
{\cal P} (w_{N-sol})= - w_{N-sol} + const.;\,\,\,\,\, {\cal P} (q_{N-sol})= q_{N-sol} + const.
\er
Therefore, one has 
\br
\label{paritys2}
{\cal P} (X)=  X,
\er
with  $X$ defined in (\ref{dpkdv1}).
Moreover, below we will assume the following boundary conditions
\br
\label{vbc1}
v(x=\pm \infty) &\rightarrow &  v_{\pm};\,\,\,\, v_x(x=\pm \infty) \rightarrow 0.\\
w(x=\pm \infty) &\rightarrow & w_{\pm};\,\,\,\, w_x(x=\pm \infty) \rightarrow 0,
\er 
with $v_{\pm}$ and $w_{\pm}$ being some constant numbers.

\section{Quasi-conservation laws: A direct approach}
\label{sec:quasi1} 

In order to find the quasi-conservation laws one can resort to either the anomalous zero-curvature \cite{npb} or  to the
Ricatti-type pseudo-potential \cite{jhep6} approaches. In these approaches the relevant quasi-conserved charges are composed by the sum of homogeneous terms defined by a suitable scaling symmetry
which was quite useful in the past in classifying the various conserved
quantities of the undeformed systems. In fact, for the pKdV model one can define the scaling
\br
\label{sc1}
v &\rightarrow & \l v,\,\,\,\, w \rightarrow \l^{-1} w\\
 t &\rightarrow &\l^{-3} t,\,\,\,\,\, x \rightarrow \l^{-1} t. \label{sc2}
\er 
So, the deformed pKdV model (\ref{dpkdv}) under the scaling transformation (\ref{sc1})-(\ref{sc2}) can be written as
\br
\label{scdpkdv}
v_t + \frac{\a}{2} v_x^2 + \, v_{xxx} &=& - \l^{-2} v_x -\epsilon_2 \frac{\a}{4} w_x v_t + \epsilon_1 (\l^{2} v_{xxt} + v_{xxx} ).
\er
Since the term $v_x$ in  (\ref{dpkdv}) can be removed by the transformation $x \rightarrow x- t$, one can conclude that the undeformed pKdV model ($\epsilon_1=\epsilon_2=0$) is invariant  under the scaling symmetry (\ref{sc1})-(\ref{sc2}), and the relevant conserved charges will inherit this property. Let us emphasize that in the developments of the quasi-integrability concept applied to deformations of the integrable models such as  SG, NLS, and KdV  \cite{jhep1, jhep2, npb20, jhep6}, the quasi-conserved charge densities resemble to the ones of the relevant integrable counterparts, such that their homogeneous terms do not contain explicitly the corresponding deformation parameters. So, one can argue that the constant asymptotic behavior of the charges is due to the quasi-integrable deformations. Therefore, the search for quasi-conserved charge densities which depend explicitly on the deformation parameters and terms with mixed scale dimensions deserve to be analyzed carefully in each case. In the case of deformations of the NLS and KdV this construction has been performed in \cite{ijmpb1,ijmpb2} and \cite{jhep6}, respectively.     

So, here we pursue the quasi-conservation laws following a direct approach starting from the deformed pKdV eqs. of motion (\ref{dpkdv})-(\ref{wv1}), which allows one to uncover the charges with mixed scaling dimension terms. So, taking the $x-$derivative of  (\ref{dpkdv}) one can write
\br
\label{quasi10}
\pa_t (v_x) + \pa_x[v_x + \frac{\a}{2} v_x^2 + v_{xxx} - X] &=& 0.
\er
This defines a first exact conservation law of the model (\ref{dpkdv}) with conserved charge
\br
\label{topo1}
q^{(1)}_X &=& \int dx \, v_x\\
&=& v_{+}-v_{-},
\er   
where  the b.c. (\ref{vbc1}) has been considered. One has that $q^{(1)}_X $ defines a topological charge. Notice that this charge is conserved even for the $X-$deformed pKdV model (\ref{dpkdv}).
By direct construction starting from the eq. (\ref{quasi10}) one can write the next quasi-conservation laws
\br
\label{quasi11}
\pa_t (v_x^2) + \pa_x[v_x^2 + \frac{2\a}{3} v_x^3 + 2 v_x v_{xxx} - v_{xx}^2 ] &=& 2 v_x X_x,\\
\label{quasi111}
\pa_t (v_x^2+ \frac{1}{2} \epsilon_1 v_{xx}^2) + \pa_x[v_x^2 + \frac{2\a}{3} v_x^3 + 2 v_x v_{xxx} - (1-\epsilon_1)v_{xx}^2 -2 v_x X] &=& \frac{1}{2} \a \epsilon_2 v_{xx} w_x v_t,\\
\label{quasi12}
\pa_t (\frac{\a}{3} v_x^3) + \pa_x[\frac{\a}{3} v_x^3 + \frac{\a^2}{4} v_x^4] &=& - \a v_x^2 v_{4x}  +\a  v_x^2  X_x,\\
\nonumber
\pa_t (\frac{\a}{3} v_x^3+ \epsilon_1 \a v_x v_{xx}^2) + \pa_x[\frac{\a}{3} v_x^3 + \frac{\a^2}{4} v_x^4] &=&(\epsilon_1-1 )\a v_x^2 v_{4x}  + \a v_{xx}\times \\
&& ( \frac{\a}{2}\epsilon_2 v_x  w_x v_t +
\epsilon_1 v_{xt} v_{xx}),\,\,\,\,\,\,\,\,\,\, \label{quasi122}\\
\pa_t (v_{xx}^2) + \pa_x[v_{xx}^2 + \a v_{xx} (v_x^2)_x - \a v_{xxx} v_x^2 + 2 v_{xx} v_{4x} - v_{xxx}^2 ] &=& -\a v_x^2 v_{4x}  + 2   v_{xx}  X_{xx},\label{quasi13}\\
\nonumber
\pa_t (v_{xx}^2 + \epsilon_1 v_{xxx}^2) + \pa_x[v_{xx}^2 + \a v_{xx} (v_x^2)_x - \a v_{xxx} v_x^2 + 2 v_{xx} v_{4x} - v_{xxx}^2 -\\
2v_{xx} X_x + 2 v_{xxx} X -2 \epsilon_1 v_{xxx} v_{xxt} - \epsilon_1 v_{xxx}^2] &=& -\a v_x^2 v_{4x}  + \frac{\a}{2} \epsilon_2    v_{4x} w_x v_t.\label{quasi133}
\er
The r.h.s. expressions in (\ref{quasi11})-(\ref{quasi133}) define the relevant anomalies and they exhibit odd parities under the symmetry transformations  (\ref{paritys1})-(\ref{paritys2}).
From the above identities one can define the following quasi-conservation laws
\br 
\frac{d}{dt} q^{(a)}_{j,X} \equiv \int_{-\infty}^{+ \infty} dx\,   \a^{(a)}_{j,X},\,\,\,\,\,\, a = 3,5;\,\,\,\,\,\,j=1,2,...4.
\er
with
\br
\label{charge31}
q^{(3)}_{1,X}  &=& \int_{-\infty}^{+ \infty} dx\,  v_x^2,\\
\label{charge32}
q^{(3)}_{2,X}  &=& \int_{-\infty}^{+ \infty} dx\,  [v_x^2+\frac{1}{2}\epsilon_1 (v_{xx})^2],  \\
\label{charge51}
q^{(5)}_{1,X}  &=& \int_{-\infty}^{+ \infty} dx\, \frac{\a}{3} v_x^3,  \\
\label{charge511}
q^{(5)}_{2,X}  &=& \int_{-\infty}^{+ \infty} dx\, [\frac{\a}{3} v_x^3 +\epsilon_1 \a v_x v_{xx}^2],  \\
\label{charge52}
q^{(5)}_{3,X}  &=& \int_{-\infty}^{+ \infty} dx\, (v_{xx})^2,\\
\label{charge53}
q^{(5)}_{4,X}  &=& \int_{-\infty}^{+ \infty} dx\, [v_{xx}^2 + \epsilon_1 v_{xxx}^2].
\er  
and 
\br
\label{ancharge31}
 \a^{(3)}_1  &=&    2 v_x X_x,\\
\label{ancharge32}
 \a^{(3)}_2    &=& \frac{\a}{2} \epsilon_2 v_{xx} w_x  v_t,\\
\label{ancharge51}
\a^{(5)}_1  &=&  [ - \a v_x^2 v_{4x}  +\a  v_x^2  X_x],\\
\label{ancharge511}
  \a^{(5)}_2   &=&(\epsilon_1-1 )\a v_x^2 v_{4x}  + \a v_{xx} ( \frac{\a}{2}\epsilon_2 v_x  w_x v_t +
\epsilon_1 v_{xt} v_{xx})\\
\label{ancharge52} 
\a^{(5)}_3  &=& [ -\a v_x^2 v_{4x}  + 2   v_{xx}  X_{xx}].\\
\label{ancharge53}
 \a^{(5)}_4  &=&  [ -\a v_x^2 v_{4x}  + \frac{\a}{2} \epsilon_2    v_{4x} w_x v_t].
\er 
Notice that in general  the charges and anomalies defined above do not exhibit homogeneous terms under the scaling symmetry   (\ref{sc1})-(\ref{sc2}). The charge $q^{(3)}_1$ in (\ref{charge31}) becomes an exact conserved charge for undeformed pKdV $(X=0)$, whereas  the charges $q^{(5)}_1$ (\ref{charge51})  and  $q^{(5)}_3$ (\ref{charge52}) define the first quasi-conserved charges even for $X=0$. Notice that $q^{(3)}_2$ becomes an exact conserved charge for $\epsilon_2=0$. Remarkably, any linear combination of the above charges define another quasi-conserved charge. In particular, one can write the next linear combination 
\br
\label{charge5d}
q^{(5),\g}_{X} & \equiv & q^{(5)}_{1,X} - \g \,  q^{(5)}_{3,X}\\
&=& \int_{-\infty}^{+ \infty} dx\, [\frac{\a}{3} v_x^3 - \g \, v_{xx}^2],
\er    
such that 
\br
\label{charge5di}
\frac{d}{dt} q^{(5),\g}_X =   \int_{-\infty}^{+ \infty} dx\,[ - \a (1- \g) v_x^2 v_{4x}  +\a  v_x^2  X_x  - 2 \g v_{xx}  X_{xx} ],
\er
with $\g$ being an arbitrary real parameter.
So, $q^{(5),\g}_X$ becomes a quasi-conserved charge of the deformed pKdV model. From (\ref{charge5d})-(\ref{charge5di}) one can see that even for undeformed pKdV ($X=0$) one has a one-parameter family of quasi-conserved charges $q^{(5),\g}_{X=0}$ parametrized by $\g$. However, for $X=0$ and  $\g=1$, one gets  
\br
\label{charge5d1}
q^{(5),\g=1}_{pKdV} & = & \int_{-\infty}^{+ \infty} dx\, [\frac{\a}{3} v_x^3 - \, v_{xx}^2],
\er 
which is an exact conserved charge of the pKdV model. 

Let us write the quasi-conserved charge $q^{(5),\g}_{X=0}$ of the undeformed pKdV model and  its relevant anomaly from (\ref{charge5di}) as
\br
\label{q5gX0}
\frac{d}{dt} q^{(5), \g}_{X=0} \equiv - \a (1- \g)  \int_{-\infty}^{+ \infty} dx\,  \a^{(5)},\,\,\,\,\,\,\,\,\,\,\,\a^{(5)} \equiv  v_x^2 v_{4x}.
\er 
Notice that for $\g=1$ one has an exact conservation law, whereas for $\g \neq 1$ this charge behaves as an asymptotically conserved  charge, i.e.
\br
\label{q5const}
q^{(5), \g}_{X=0}(t = +\infty)  =  q^{(5), \g}_{X=0} (t = -\infty),
\er
provided that
\br
\label{ano5}
\int_{t = - \infty }^{t = + \infty} dt\, \int_{-\infty}^{+ \infty} dx\,  \a^{(5)} =0.
\er
 
Note that $X$ in the the r.h.s. of (\ref{dpkdv}) contains  a non-local term $w_x$, since this can be defined from (\ref{wv1}) in terms of the pKdV field $v$ as $w_x = \int dt \, v_x$. It is interesting to search for quasi-conservation  laws with charges incorporating this type of non-local terms. So, from (\ref{dpkdv}) one can write 
\br
\nonumber
\pa_t \Big[ v_x X -v_x^2 - \frac{\a}{3} v_x^3 + v_{xx}^2 -v_x v_t \Big] - \pa_x\Big[v_x v_{xxt} - X(v_x + \frac{\a}{2} v_x^2 +v_{xxx}) + \\ X^2-v_t^2 - v_t v_x -v_t v_{xxx} + v_{xt} v_{xx} \Big]  = - \frac{\a}{2} v_t (v_x^2)_x
\label{nonloc1}
\er
with non-local charge 
\br
\label{nonlocharge1}
\hat{q} \equiv \int dx\,[v_x X -v_x^2 - \frac{\a}{3} v_x^3 + v_{xx}^2 -v_x v_t ]  
\er
and anomaly density
\br
\label{nonlocano1}
 \hat{\a}\equiv - \frac{\a}{2} v_t (v_x^2)_x.
\er 
Below we will perform numerical simulations of some of the above anomalies  for the collision of 2-kink solitons of the (deformed) pKdV  model.  We will numerically verify the quasi-conservation law (\ref{quasi133}) and the relevant anomaly density $\a^{(5)}_4(x,t)$ in (\ref{ancharge53}) in the Fig. 1. Likewise, we  will examine the behavior of the anomaly (\ref{nonlocano1}) in the Fig. 2. 

Next, let us consider the equivalent equation of the model (\ref{dpkdv}) in the $q-$field parametrization (\ref{eqq}) and write the following quasi-conservation law
\br
\nonumber
\pa_t[\frac{1}{2} q_{t}^2 -  \epsilon_2  q_{xx} q_{t}^2 +  \frac{1}{2}\epsilon_1 q_{xt}^2] + \pa_x[ \frac{1}{2} q_{t}^2 - 4 q_{xt} q_t^2 + q_{xxt} q_{t} - \frac{1}{2} (1-\epsilon_1)q_{xt}^2 - \epsilon_1  q_{xtt} q_{t} -  \epsilon_1 q_{xxt} q_t] \\ = - 2(2 + \epsilon_2) q_{xxt} q_t^2 - 4 q_{xt}^2 q_t . \label{eqq1}
\er
Then, one can define the charge and anomaly, respectively, as
\br
\label{calQ11}
{\cal Q}_1 &=& \int dx [\frac{1}{2} q_{t}^2 -  \epsilon_2  q_{xx} q_{t}^2 +  \frac{1}{2}\epsilon_1 q_{xt}^2],\\
{\cal A}_1&=& \int dx [ - 2(2 + \epsilon_2) q_{xxt} q_t^2 - 4 q_{xt}^2 q_t]. 
\er
Note that the charge ${\cal Q}_1$ diverges when evaluated on the vacuum solution $q_{clb}$ (\ref{clb}), due to the contribution of the term $\frac{1}{2} q_{t}^2$. Therefore, the charge  ${\cal Q}_1$ (\ref{calQ11}) for the composite field $q \sim `kink' + q_{clb}$ must be renormalized by subtructing the contribution (which is infinite) of the continuous linear background $q_{clb}$ in  (\ref{clb}), i.e. the kink charge ${\cal Q}_{1\,kink}$ must be renormalized. This is reminiscent of the procedure performed in the defocusing NLS model in which the momentum associated to the dark soliton itself has been renormalized \cite{jhep4}.

\section{Tau function and direct methods to construct solitons}
\label{sec:tau1}

Next we use the tau function method in order to obtain the 1-soliton and 2-soliton solutions of the model (\ref{dpkdv}) for some particular cases of the deformations parameters $\{\epsilon_1,\epsilon_2\}$.
 
Let us consider
\br
\label{qtau11}
q_s(x, t) =   
\b \log{\tau(x,t)},
\er
where $\tau$ defines  a tau function.  
Next, replacing  (\ref{qtau11}) into (\ref{eqq}) one gets the next equation
\br
&& 
 \b  \tau ^3 \left[-\epsilon_1 \tau_{xxtt}+(1-\epsilon_1) \tau_{xxxt}+ \tau_{tt}+ \tau_{xt} \right]\nonumber\\
&-& \b \tau ^2 \left[(2 \beta {\epsilon}_2 -\epsilon_1) \tau_{tt} \tau_{xx}+ 2(2\beta-\epsilon_1)
   \tau_{xt}^2+\tau_{t} \left(-2 \epsilon_1 \tau_{xxt}+(1-\epsilon_1) \tau_{xxx}+ \tau_{x}\right)
  \right. \nonumber\\
   &-&2 \left.\epsilon_1
   \tau_{x} \tau_{xtt}+3 (1-\epsilon_1) \tau_{xt} \tau_{xx}+3 (1-\epsilon_1) 
   \tau_{x} \tau_{xxt}+ \tau_{t}^2+3 (1-\epsilon)\tau_{x} \tau_{xxt}\right]
   \nonumber\\
   &+&2\b  \tau  \left[(\beta {\epsilon}_2-\epsilon_1) \tau_{xx}
   \tau_{t}^2+(\beta {\epsilon}_2-\epsilon_1) \tau_{tt} \tau_{x}^2
   \right. \nonumber\\
  & +& \left. \tau_{x} \tau_{t} \left(4 (\beta-\epsilon_1) \tau_{xt}-3 (\epsilon_1-1) \tau_{xx}\right)
   -3 (\epsilon_1-1) \tau_{x}^2 \tau_{xt}\right]
   \nonumber\\
   &-& 2 \tau_{t} \tau_{x}^2
   \left[(\beta ({\epsilon}_2+2)-3 \epsilon_1) \tau_{t}-3 (\epsilon_1-1) \tau_{x}\right] = 0.
   \label{taueq1}
   \er 

{\bf 1-soliton solution of deformed pKdV}

The tau function method furnishes the first type  of kink solution of (\ref{eqq}) for arbitrary values of the set $\{\epsilon_1,\epsilon_2\}$. So, substituting the tau function $\tau = 1 + e^{\G_1}$ into (\ref{taueq1}), and taking into account (\ref{qtau11}), one has  
\br
q_{1} &=& \frac{3}{(2+\epsilon_2)\left(1+(1-\epsilon_1) k^2\right)}   \log{[1+e^{\G_1}]} \\
&=& \frac{3}{(2+\epsilon_2)\left(1+(1-\epsilon_1) k^2\right)}  \Big\{ \log{2}+ \frac{\G_1}{2}+ \log{\cosh{(\frac{\G_1}{2})}}  \Big\},
\er
with
\br
\G_1 = k_1 x - w_1 t + \d; \,\,\,\,\,\, w_1 = \frac{k_1+ (1-\epsilon_1) k_1^3}{1- \epsilon_1 k_1^2}.
\er
So, the eq.  (\ref{wxvt}) provides  the first type of 1-soliton solution for the pKdV field $v$
\br
\label{v11}
v_1= \frac{12}{\a} \frac{k_1}{(2+\epsilon_2)\( 1-\epsilon_1 k_1^2\)} \mbox{Tanh}\Big[\frac{1}{2} \(k_1 x - w_1 t + \d\)\Big].
\er
 
{\bf 2-soliton solution of deformed pKdV: the case $\epsilon_1 =\epsilon_2 =1$}
 
The pKdV $2-$soliton solution follows similar construction as in the case of the modified KdV for the particular case $\epsilon_1 =\epsilon_2 =1$. The field $q$ takes the form \cite{npb, gibbon}
\br
\label{qq22}
q &=& \log{\Big[1+ e^{\G_1}+e^{\G_2}+ A_{12} e^{\G_1} e^{\G_2}\Big]},\,\,\,\, \G_i = k_i x - w_i t + \delta_i,\,\,\,w_i= \frac{k_i}{1-k_i^2},\,\,i=1,2.\\
A_{12}&=& -\frac{(w_1 - w_2)^2 (k_1 - k_2)^2 + (w_1 - w_2)(k_1 - k_2) - (w_1 - w_2)^2}{(w_1 + w_2)^2 (k_1 + k_2)^2 + (w_1 + w_2)(k_1 + k_2) - (w_1 + w_2)^2}.
\er
In order to exhibit the parity symmetry we follow the approach in \cite{jhep6} developed for the modified KdV. So, in order to perform the transformation (\ref{parity1}) and check the space-time parity inversion symmetry of the 2-soliton solution we will derive a new expression for $q$ in (\ref{qq22}), such that  $v_{2-sol}$ in (\ref{wxvt}) becomes a manifestly ${\cal P}$ invariant function. So, let us define a new parameter $\D$, as $A_{12} = e^{\D}$, and
\br
\label{etas1}
 \G_j &=& k_j \widetilde{x} - w_j \widetilde{t} + \eta_{0j} - \frac{\D}{2} \equiv \eta_j - \frac{\D}{2}, \,\,\,\,\, j=1,2  
\er
where 
\br \label{deltas1}
\delta_j  = -k_j x_{\D} + w_j t_{\D} + \eta_{0j} - \frac{\D}{2}, \,\,\,j=1,2.
\er
Therefore, $q$ can be rewritten as 
\br
q = \log{\Big[2 e^{-\D/4} \, e^{(\eta_1+\eta_2)/2} \(e^{\D/4} \cosh{(\frac{\eta_1+\eta_2}{2})} + e^{-\D/4}  \cosh{(\frac{\eta_1-\eta_2}{2})}  \)\Big]}.
\er
So, using (\ref{wxvt}) one has 
\br
\label{u2ee}
v_{2-sol} &=&\frac{4}{\a} (w_1+w_2) - \frac{8}{\a} \pa_t   \log{\Big[ e^{\D/4} \cosh{(\frac{\eta_1+\eta_2}{2})} + e^{-\D/4}  \cosh{(\frac{\eta_1-\eta_2}{2})} \Big]}.
\er
Therefore, the  2-soliton  
\br
\label{usolxt}
{\bf v}_{2-sol} = v_{2-sol} \Big|_{\eta_{01}=\eta_{02}=0},
\er
satisfies the symmetry
\br
\label{P2sol}
{\cal P}({\bf v}_{2-sol})= -{\bf v}_{2-sol} + const.
\er 
Moreover, using (\ref{wxvt}) and following similar steps as above one can show
\br
\label{parity3}
{\cal P}({\bf w}_{2-sol}) = - {\bf w}_{2-sol} + const.
 \er 
Due to the condition $\eta_{01}=\eta_{02}=0$, one can get the next relationships for the coordinates $(x_{\D},t_{\D})$
\br
x_{\D} &\equiv & \frac{w_2 \widetilde{\theta}_1 - w_1 \widetilde{\theta}_2}{k_2 w_1 - k_1 w_2}\\
t_{\D} & \equiv & \frac{k_2 \widetilde{\theta}_1 - k_1 \widetilde{\theta}_2}{k_2 w_1 - k_1 w_2},\,\,\,\,\,\widetilde{\theta}_j \equiv \frac{\D}{2} + \delta_j,\,\,\,j=1,2.  
\er  

\subsection{Tau function approach and ${\cal P}$ symmetric N-solitons of the usual pKdV}
  
In order to find the N-soliton solutions of the pKdV model satisfying the symmetry properties (\ref{parity1}) and (\ref{paritys1}) we closely follow the construction performed for the KdV model in \cite{alice, jhep6}.  So, we will construct a general $N-$soliton solution possessing the space-time parity symmetry (\ref{parity1})-(\ref{paritys1}), for any shifted point and delayed time ($x_\Delta,\,t_\Delta$) in space-time. 

The usual pKdV equation  of motion  is defined by setting $\epsilon_1=\epsilon_2=0$ and making the transformation $x\rightarrow x-t$ in (\ref{dpkdv}); so, one has
\br
\label{pkdv1}
v_t +\frac{\a}{2} v_x^2 + \, v_{xxx} = 0.
\er
The Hirota's tau function for the eq. (\ref{pkdv1}) is introduced as
\br
\label{vlog}
 v  = \frac{12}{\alpha} \pa_{x} \log{\tau}. 
\er 
The Hirota bi-linear equation of (\ref{pkdv1}), as well as multi-solitons, lumps, and breather wave-solutions have recently been presented in \cite{alhami}. Since the pKdV field $v$ and the KdV field $u$ can be related by $u = \pa_x v$, one can take the same  tau functions for the both models in order to construct their soliton solutions. So, let us assume the next tau function for the $N-$soliton solution \cite{hirota1}
\br
\label{Nsol}
\tau_N = \sum_\mu \exp\(\sum_{j=1}^{N} \mu_j \G_{j} + \sum_{1\leq j< l}^{N} \mu_j \mu_l \theta_{jl}\) 
\er
where the $\mu-$summation is undertaken in all the permutations of $\mu_i = 0,1$, for $i=1,2,...N$, and
\br
\G_j = k_i x - w_i t + \xi_{0j},\,\,\,\,e^{\theta_{ij}} = \(\frac{k_i - k_j}{k_i + k_j}\)^2,\,\,\,\,w_i =  k_i^3.
\er  
The $\xi_{0j}'s$ are arbitrary constants related to the space-time translation invariance of the pKdV equation, such that  each $j-$soliton component of the $N-$soliton can be located anywhere at $\xi_{0j}$. A subset of solutions possessing the space-time symmetry (\ref{parity1})-(\ref{paritys1}) is achieved by making a particular choice of the set of parameters  $\xi_{0j}$, such that the space-time translation symmetry of the solution (\ref{Nsol}) is broken. So, let us consider \cite{jhep6}
\br
\G_j = k_j (x-x_{\Delta}) - w_j (t-t_\D) + \eta_{0j} - \frac{1}{2} \sum_{i=1}^{j-1} \theta_{ij} -  \frac{1}{2} \sum_{i=j+1}^{N} \theta_{ji} \equiv \eta_j -  \frac{1}{2} \sum_{i=1}^{j-1} \theta_{ij} -  \frac{1}{2} \sum_{i=j+1}^{N} \theta_{ji}. 
\er
With these redefinitions and the tau function  (\ref{Nsol}) one can write the $N-$soliton solution in the equivalent form
\br
\label{Nsol1}
 v_N  =  -\frac{6}{\alpha} (\sum_{j }k_j ) + \frac{12}{\alpha} \pa_{x} \Big[\log{ \sum_{\nu} K_\nu \cosh{\(\frac{1}{2} \sum_{j=1}^{N} \nu_j \eta_j \)}}   \Big],
\er
where the summation in $\nu$ involves all the permutations of $\nu_i = 1, -1$, $i=1,2,...N$, and $K_\nu = \Pi_{i>j} (k_i - \nu_i \nu_j k_j)$.

Then, from (\ref{Nsol1}) let us define the $N-$soliton solution with space-time translation symmetry broken as 
\br
\label{usym}
{\bf v}_N = v_N|_{\eta_{0j}=0}. 
\er
It is a matter of direct verification  that this solution ${\bf v}_N$ will exhibit the symmetry (\ref{parity1})-(\ref{paritys1}), i.e.
\br
\label{sym0}
{\cal P} ({\bf v}_N)= - {\bf v}_N + const.
\er
Next, we present the above constructions for the cases $N=1, 2, 3$, and describe the main properties of the corresponding solitons. 

{\bf Case $N=1$}. One has  $\tau_1 = 1 + e^{\G_1}$, which  can be written as 
\br
\tau_1 = 2 e^{-\frac{\eta_1}{2}}\Big[\cosh{\frac{\eta_1}{2}}\Big],\,\,\,\,\eta_1 = k_1 (x-x_{\Delta}) - w_1 (t-t_\D) + \eta_{01}.
\er
Next, one has
\br
\label{tau11}
\log{(\tau_1)} = \log{2} - \frac{\eta_1}{2} + \log{\Big[\cosh{\frac{\eta_1}{2}}\Big]}.
\er
Therefore, inserting (\ref{tau11}) into (\ref{vlog}) one can write
\br
{\bf v}_1 = -\frac{6 k_1}{\alpha} + \frac{12}{\alpha} \pa_x  \log{\Big[\cosh{\frac{\eta_1}{2}}\Big]}|_{\eta_{01}=0}.
\er 
This expression satisfies  the symmetry transformation  ${\cal P}({\bf v}_1)= -{\bf v}_1 + const.$ This soliton can be written as
 \br
{\bf v}_1 = -\frac{6 k_1}{\alpha}\,\Big\{ 1- \mbox{Tanh}\Big[\frac{k_1 (x-x_\Delta) - w_1 (t-t_\D)}{2}\Big]\Big\}.
\er
This is the topological 1-kink solution of the pKdV model (\ref{pkdv1}). 

{\bf Case $N=2$}. The tau function becomes 
\br
\tau_2 = 1 + e^{\G_{1}} + e^{\G_2} +  e^{\G_{1} + \G_2 + \theta_{12}}, 
\er
which, following  the above construction, can be rewritten as 
\br
\tau_2 &=& \frac{2}{k_1-k_2} e^{(\eta_1+\eta_2)/2}\Big[ (k_1-k_2) \cosh{\(\frac{\eta_1+\eta_2}{2}\)} + (k_1+k_2)  \cosh{\(\frac{\eta_1-\eta_2}{2}\)} \Big].\\
\eta_i &=& k_i (x-x_\D) - w_i (t-t_\D) + \eta_{0i},\,\,\,\,i=1,2.
\er    
Then, inserting (\ref{tau11}) into (\ref{vlog}) and taking into account (\ref{Nsol1}) and (\ref{usym}) it is straightforward to construct a ${\cal P}$ invariant 2-soliton as 
\br
{\bf v}_2  = -\frac{6}{\alpha} (k_1 + k_2) +\frac{12}{\alpha} \pa_{x} \log{ \Big[ (k_1-k_2) \cosh{\(\frac{\eta_1+\eta_2}{2}\)} + (k_1+k_2)  \cosh{\(\frac{\eta_1-\eta_2}{2}\)} \Big]}\Big|_{\eta_{01}=\eta_{02}=0}.
\er
Thus, this pKdV 2-soliton solution transforms under the parity transformation as ${\cal P}({\bf v}_2)= - {\bf v}_2 + const.$   
  
{\bf Case $N=3$}.  It can be constructed following similarly steps. So, the tau function $\tau_3$ becomes
\br
\tau_3 =  1 + e^{\G_{1}} + e^{\G_2} +  e^{\G_{3}} + e^{\G_1 + \G_{2} + \theta_{12}} + e^{\G_1 + \G_{3} + \theta_{13}} + e^{\G_2 + \G_{3} + \theta_{23}} + e^{\G_1 + \G_{2} + \G_3 + \theta_{12}+ \theta_{13}+\theta_{23}}, 
\er 
which can be rewritten as 
\br
\tau_3 &=& \frac{2 e^{(\eta_1+\eta_2+\eta_3)/2}}{(k_1-k_2)(k_1-k_3)(k_2-k_3)}\Big[ C(x,t)\Big]\\ \nonumber
    C(x,t)  &\equiv& (k_1-k_2)(k_1-k_3)(k_2-k_3) \cosh{[(\eta_1+\eta_2+\eta_3)/2]} + \nonumber\\
&&(k_1+k_2)(k_1+k_3)(k_2-k_3) \cosh{[(-\eta_1+\eta_2+\eta_3)/2]} + \nonumber\\
&&
(k_1+k_2)(k_1-k_3)(k_2+k_3) \cosh{[(\eta_1-\eta_2+\eta_3)/2]}+\nonumber\\
&& (k_1-k_2)(k_1+k_3)(k_2+k_3) \cosh{[(\eta_1+\eta_2-\eta_3)/2]},\nonumber\\
\eta_{i}&=& k_i (x-x_\D) - w_i (t-t_\D) + \eta_{0i},\,\,\,\,i=1,2,3. \nonumber  
\er
Similarly, it is straightforward to construct a ${\cal P}$ invariant 3-soliton as 
\br
{\bf v}_3  =  -\frac{6}{\alpha} (k_1 + k_2+k_3)  + \frac{12}{\alpha} \pa_{x} \log{ C(x,t)} \Big|_{\eta_{01}=\eta_{02}=\eta_{03}=0}.
\er
Clearly, this pKdV 3-soliton solution transforms as ${\cal P}({\bf v}_3 ) = - {\bf v}_3 + const.$ 

So, the  ${\cal P}$ invariant N-soliton solutions would allow us to show that the  $(x,t)-$integrated anomalies of the integrable pKdV model vanish. In particular, for the quasi-conserved charges $q^{(5)}_1$ (\ref{charge51})  and  $q^{(5)}_3$ (\ref{charge52}) defined for $X=0$, one can show the vanishing of the $(x,t)-$integrated anomaly $-\a v_x^2 v_{4x}$ (\ref{ancharge51}) (or (\ref{ancharge52}))  using the parity ${\cal P}$ argument. 

\subsection{A direct method  and a second type of 1-kink solutions}
\label{1soli}
 Next we present a direct method in order to find a second type of 1-soliton solution for any set $\{\epsilon_1,\epsilon_2\}$, which generalizes the one obtained above through the tau function method. In fact, a direct method provides a general 1-soliton solution of  (\ref{eqq}) by assuming the form
\br
\label{qii}
q_{I} &=& q_0 \Big\{\log{ \cosh{[\frac{\zeta}{2 a}] } }+ b \, \zeta + c \Big\},\,\,\,\,\,\zeta =  k x - w_1 t + \d.
\er
Note that we have incorporated the term $q_0 (b \, \zeta + c)$ corresponding to the continuous linear background (\ref{clb}). 
A direct substitution of $q_{I}$ into (\ref{eqq})  provides the relationships
\br
\label{disp}
w_1 = \frac{a^2 k + (1-\epsilon_1) k^3}{a^2- \epsilon_1 k^2};\,\,\,\,\, q_0 =\frac{3 a^2}{(a^2+(1-\epsilon_1) k^2)(2 + \epsilon_2)},
\er
such that $a,\,b$ and $c$ are arbitrary real parameters. So, through (\ref{wxvt}) one has the second type of 1-soliton solution for $v$
\br 
\label{solgeral}
v_{I}= -q_o b w_1 +\frac{12 a }{\a}\frac{k}{( a^2- \epsilon_1 k^2 ) (2+ \epsilon_2)} \mbox{Tanh}\Big[\frac{1}{2 a} (k x - w_1 t + \d)\Big].
\er
This is a new general form of a 1-kink soliton plus a constant background which can not be found by the tau function method. Clearly, the two types of solutions $v_1$ in (\ref{v11}) and the kink sector of $v_I$ in (\ref{solgeral}) become the same for $a^2 =1$ and for arbitrary values of the set $\{\epsilon_1, \epsilon_2\}$. However, the solution $v_I$ describes a family of kinks parametrized by an additional free parameter $a$, which can be related to the width of the kink $\sim \frac{2a}{k}$. As the initial condition of our numerical simulations for 2-kink collision below we will consider  the superposition of two solitons of the general type (\ref{solgeral}) located some distance apart.  

\section{Analytical quasi-integrability of the pmRLW theory ($\epsilon_1=\epsilon_2=1$)}
\label{sec:dpkdv11}

Let us consider a sub-model of (\ref{dpkdv}) defined by the special case $\epsilon_1=\epsilon_2=1$. So, one has the sub-model  
\br
v_t + v_x +\frac{\a}{2} v_x^2 + \frac{\a}{4} w_x v_t  - v_{xxt} =0. \label{dpkdve1e2}
\er
This model can be considered as the potential modified regularized long wave model (pmRLW). The non-integrable mRLW model has been discussed before, see  \cite{jhep6} and references therein. The analytical form of a 2-soliton solution of (\ref{dpkdve1e2}) and its ${\cal P}$ invariant representation was obtained in (\ref{usolxt})-(\ref{P2sol}).

So, taking into account that the 2-soliton (\ref{usolxt}) satisfies the parity symmetry (\ref{P2sol}), and the symmetry of the auxiliary field $w$ in  (\ref{parity3}), one can show analytically the vanishing of the anomalies belonging to the various quasi-conservation laws. Thus, it is an analytical proof of the quasi-integrability of the pmRLW theory (\ref{dpkdve1e2}), i.e. the deformed pKdV for $\epsilon_1=\epsilon_2=1$. Notice that similar argument has been used in order to present this proof for the KdV-like quasi-conservation laws in \cite{jhep6}. Here, we are generalizing this proof for the new quasi-conservation laws presented above. Then, it is 
worth to mention that this adds a new strong result on the analytical proof, not only numerical, of the quasi-integrability of a (non-integrable) theory, discussed in \cite{npb, jhep6}. 

\section{Numerical treatment of the anomalies}
\label{sec:num}

Here we numerically  simulate the vanishing of some representative anomalies of the deformed pKdV model, which have been discussed above through the symmetry arguments to advance the, so far,  only plausible explanation for the anomalous charges in the soliton collision of the deformations of the integrable models, such as the SG \cite{npb20}, NLS \cite{ijmpb1, ijmpb2} and KdV  \cite{jhep6} models, respectively. The above properties will be qualitatively reproduced in our numerical simulations of the 2-kink interaction of the deformed pKdV model, for a variety of soliton configurations and a wide range of values of the set of deformation parameters $\{\epsilon_1, \epsilon_2\}$.  
 
We have presented in (\ref{solgeral}) a general one kink-type solution of the model for any set of the deformation parameters $\{\epsilon_1, \epsilon_2\}$, and it exhibits a general dispersion relation (\ref{disp}). We will use two of this exact one soliton solution located far apart as the initial condition in order to simulate two-soliton collision for the deformed model. Our initial condition must be a field configuration which reduces  the emission of radiation in order to accurately simulate the various quantities, such as the charges and anomalies. 

Since an analytical solution for 2-soliton for any values of $\epsilon_1$ and $\epsilon_2$ of (\ref{dpkdv}) is not known; as mentioned above, we will take as an initial condition the superposition of two solitons of the general type (\ref{solgeral}) located some distance apart. So, let us consider a linear superposition of two expressions of type (\ref{qii}) 
\br
\label{q2s}
q_{2s}(x, t) &=& q_1 \Big\{\log{ \cosh{[\frac{\zeta_1}{2 a_1}] } }+ b_1 \zeta_1 + c_1 \Big\} +  q_2 \Big\{\log{ \cosh{[\frac{\zeta_2}{2 a_2}] } }+ b_2 \zeta_2 + c_2 \Big\} ,\\
w_j &=& \frac{a_j^2 k_j + (1-\epsilon_1) k_j^3}{a_j^2- \epsilon_1 k_j^2};\,\,\,\,\, q_j =\frac{3 a_j^2}{[a_j^2+(1-\epsilon_1) k_j^2](2 + \epsilon_2)};\,\,\,\zeta_j =  k_j x - w_j t + \d_j,\,\,\,\, j =1,2;
\er
such that $a_j,\,b_j$ and $c_j, \, j=1,2,$ are arbitrary real parameters. The expression of each component of the field $q_{2s}(x,t)$ is presented in (\ref{qii}). We closely follow  the numerical techniques presented in the Appendix of \cite{jhep6}.

We plot the function $q_{2s}(x, t_i)$  (upper green line in top left panel of Fig.1), and the kink $v_{2s}(x, t_i)$ (bottom green line in the top left panel of Fig.1) for initial time $t_i = 0$. The field $v_{2s}(x, t_i)$ (green) represents the initial configuration of our numerical solution of the model (\ref{dpkdv}) for two-soliton collision. Notice that the auxiliary field $q_{2s}(x, t_i)$ (green) undergoes significant changes only around the kink regions from an approximately linear behavior in regions far away from the kinks; so, realizing the continuous linear background solution in (\ref{clb}) as the vacuum of the theory. Whereas, the field $v_{2s}(x, t_i)$ behaves as a two-kink-like function and approaches constant values asymptotically for $x \rightarrow \pm \infty$. These patterns and properties will be useful when imposing the relevant initial and boundary conditions of our numerical simulations. So, in the bottom lines of the top left panel of Fig. 1 we plot the field configurations of the 2-soliton of the pKdV model. One has the kink $v_{2sol}$ (bottom lines) and the field $q_{2sol}$ (upper lines) for three successive times, i.e. the initial (green), collision (orange) and final (magenta) times, respectively,  for $\epsilon_1= 1.2, \epsilon_2 = 0.9$.. 
 
We examine numerically the quasi-conservation law (\ref{quasi133}) and the relevant anomaly density $\a^{(5)}_4(x,t)$ in (\ref{ancharge53}). The Fig.1 (top right) shows the anomaly density $\a^{(5)}_4(x,t)$ for three successive times, initial (green), collision (orange) and final (magenta), respectively. The bottom figures show the plots  $(\int_{-L}^{+ L} dx\, \a^{(5)}_4(x,t))$ and $(\int_{t = t_i }^{t } dt'\, \int_{-L}^{+ L} dx'\,  \a^{(5)}_4(x',t')$.  Notice the vanishing of the $(x,t)-$integrated anomaly at the order of $\approx 10^{-3}$. So, one can regard $q^{(5)}_{4, X}$ in (\ref{charge53}) as an asymptotically conserved charge within numerical accuracy, for the 2-kink collision.

\begin{figure}
\centering
\label{fig1}
\includegraphics[width=2cm,scale=6, angle=0,height=6cm]{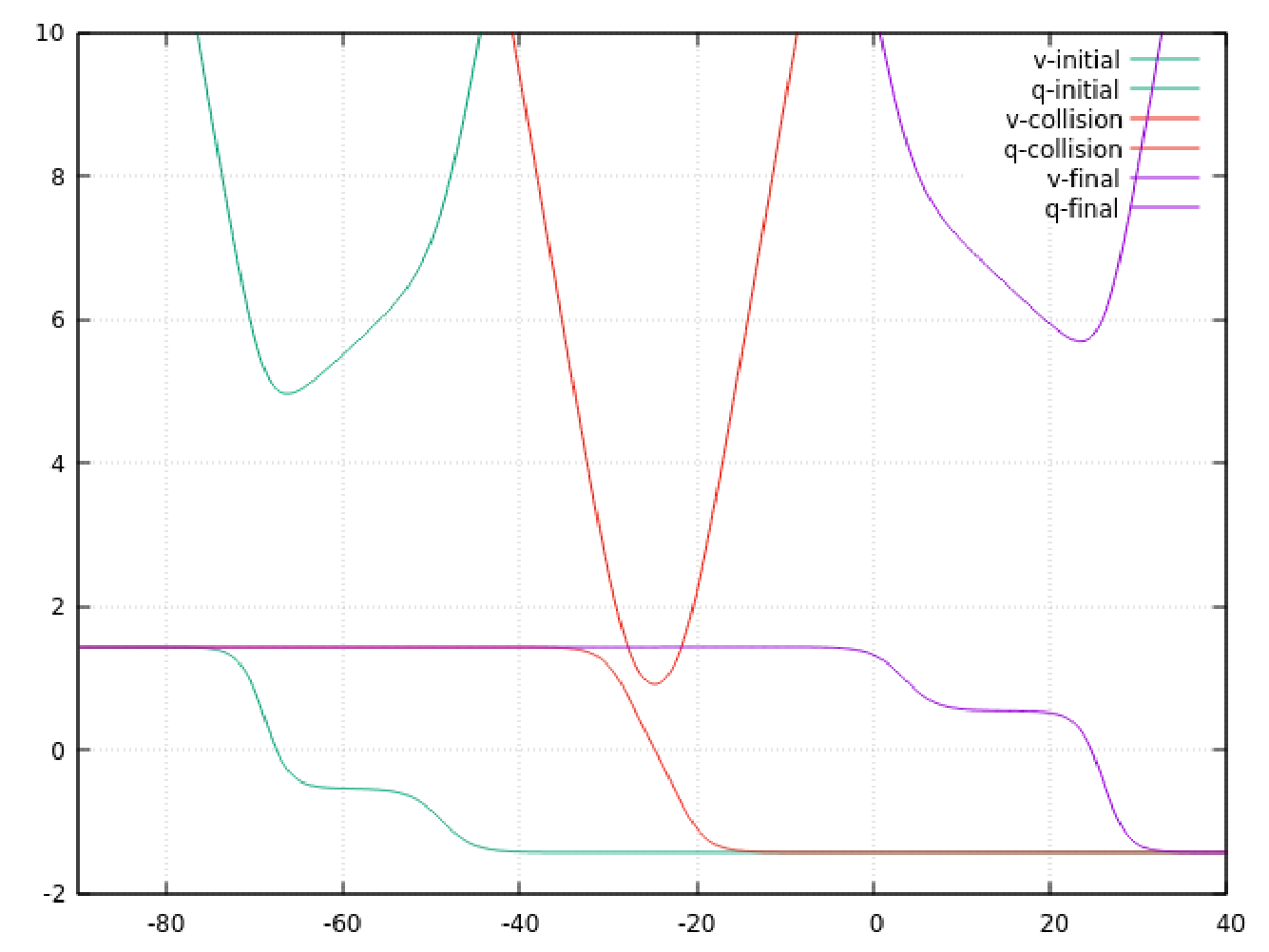}
\includegraphics[width=2cm,scale=6, angle=0,height=6cm]{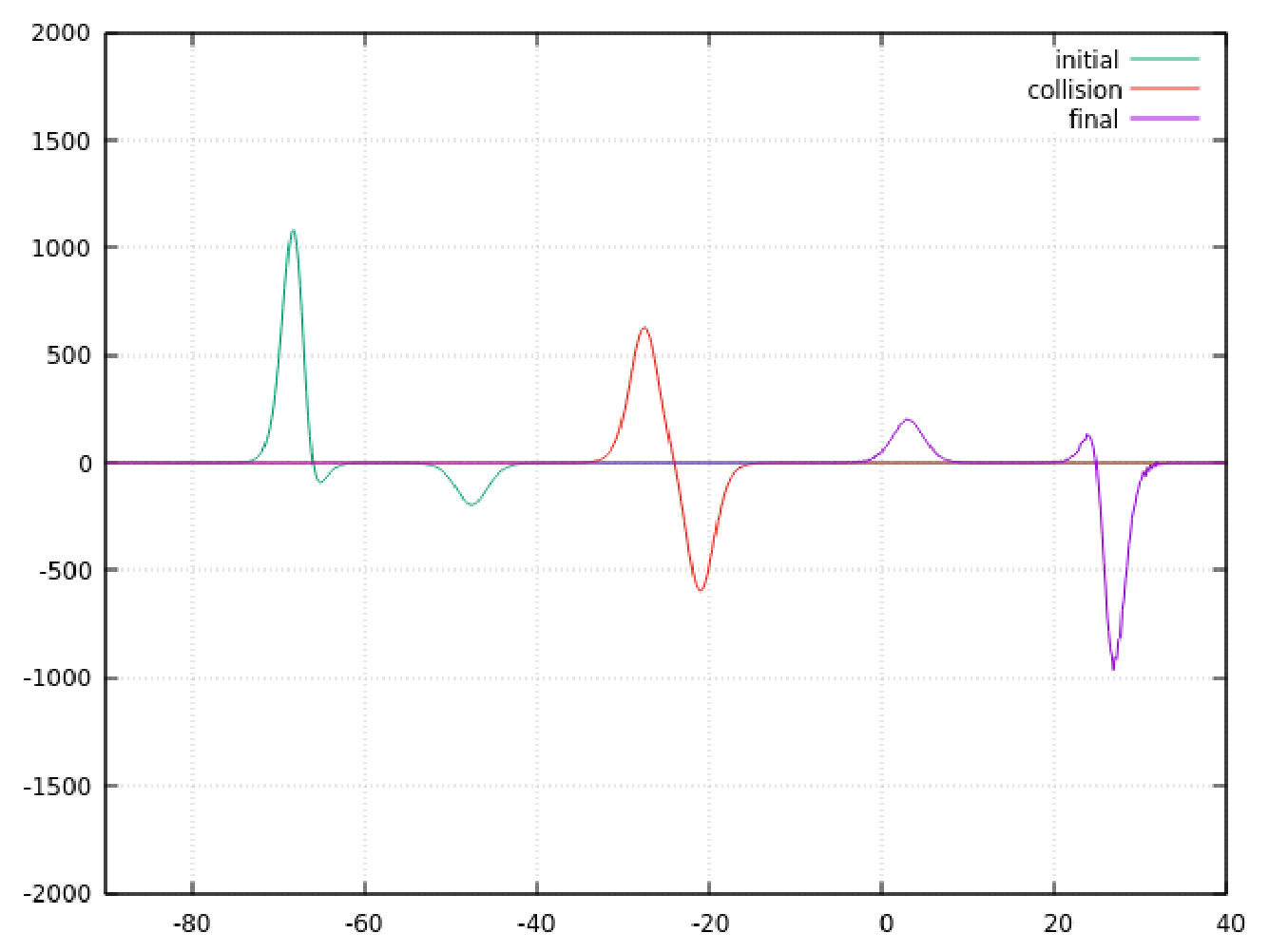} 
\includegraphics[width=2cm,scale=6, angle=0,height=6cm]{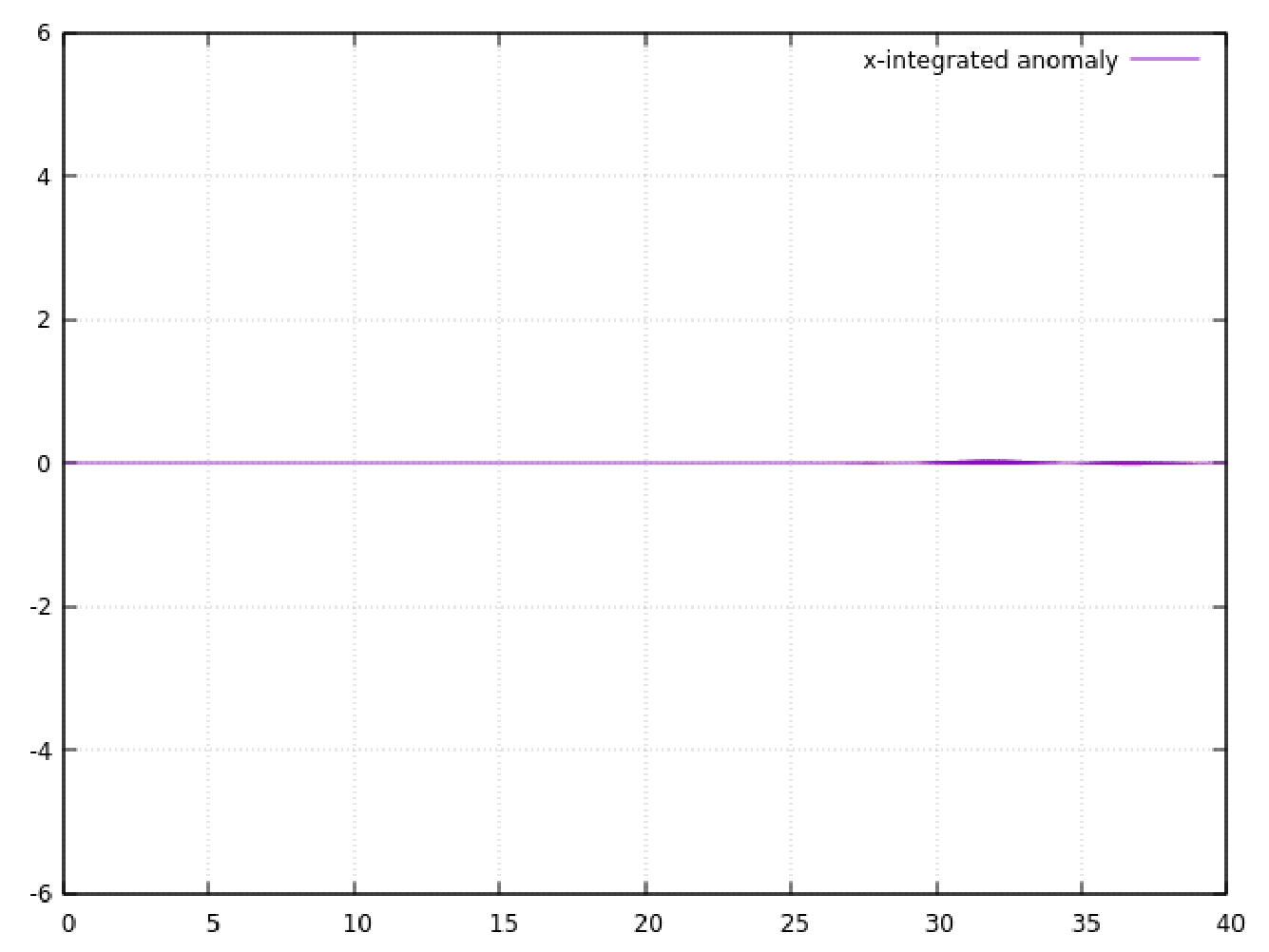}
\includegraphics[width=2cm,scale=6, angle=0,height=6cm]{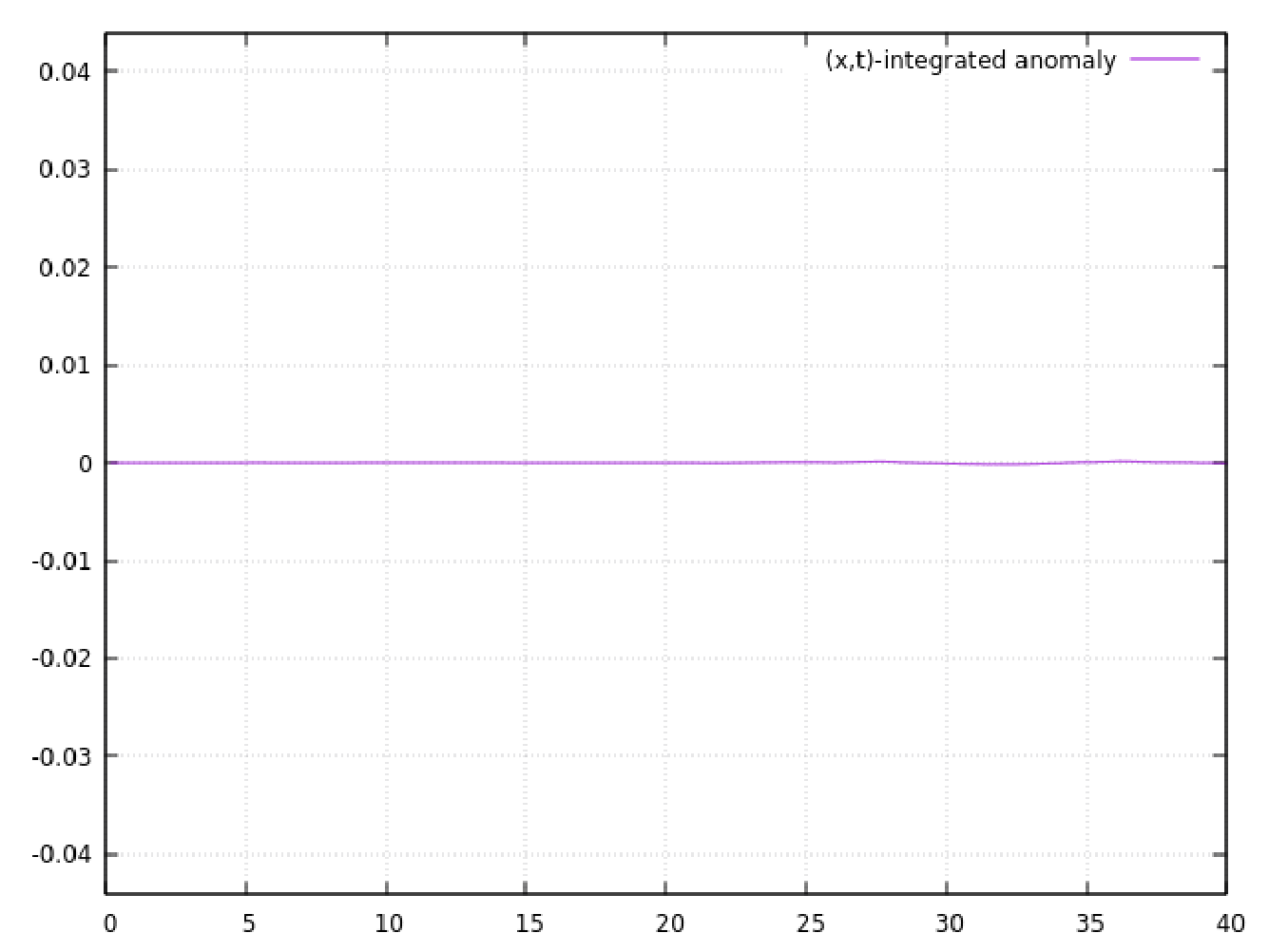}
\begin{center}
\parbox{6in}{\caption{(color online) Numerically simulated two-kink collision, $v$ field of (\ref{dpkdv}) (bottom of top left figures), and the field $q$ of (\ref{eqq})  (upper of top left figures). The top Figs. show three successive times, initial (green), collision (orange) and final (magenta), respectively, for $\epsilon_2= 0.9, \epsilon_1= 1.2$. The top right Fig. shows the anomaly density $\a^{(5)}_4(x,t)$ in (\ref{ancharge53}) for three successive times, initial, collision and final, respectively. The bottom left Fig. shows the $x-$integrated anomaly and the bottom right shows the $(x,t)-$integrated anomaly. Notice that the right bottom figure vanishes within numerically accuracy.}}
\end{center}
\end{figure}
   
Next, we examine numerically the non-local quasi-conservation law (\ref{nonloc1}) and the relevant anomaly density $\hat{\a}$ in (\ref{nonlocano1}). The plot in  Fig.2 (top Fig.) shows the anomaly density $\hat{\a}$ for three successive times, initial (green), collision (orange) and final (magenta), respectively. The middle and bottom  figures show the plots  $(\int_{-L}^{+ L} dx\, \hat{\a}(x,t))$ and $(\int_{t = t_i }^{t } dt'\, \int_{-L}^{+ L} dx'\,  \hat{\a}(x',t')$.  Notice the vanishing of the $(x,t)-$integrated anomaly at the order of $\approx 10^{-9}$. So, one can regard $\hat{q}$ in (\ref{nonlocharge1}) as a non-local charge which is asymptotically conserved within numerical accuracy, for the 2-kink collision.

\begin{figure}
\centering
\label{fig2}
\includegraphics[width=1cm,scale=2, angle=0,height=6cm]{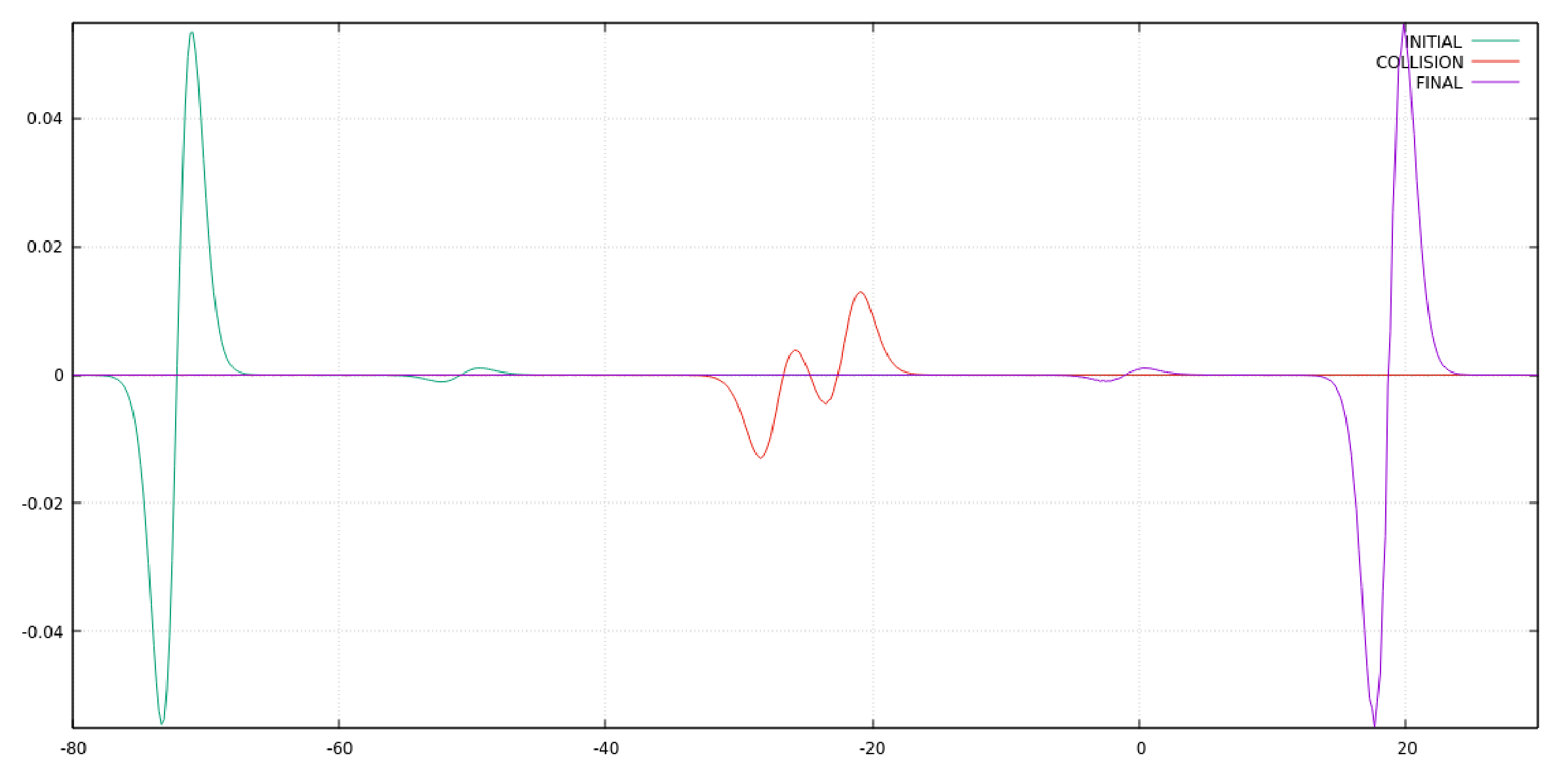} 
\includegraphics[width=1cm,scale=2, angle=0,height=6cm]{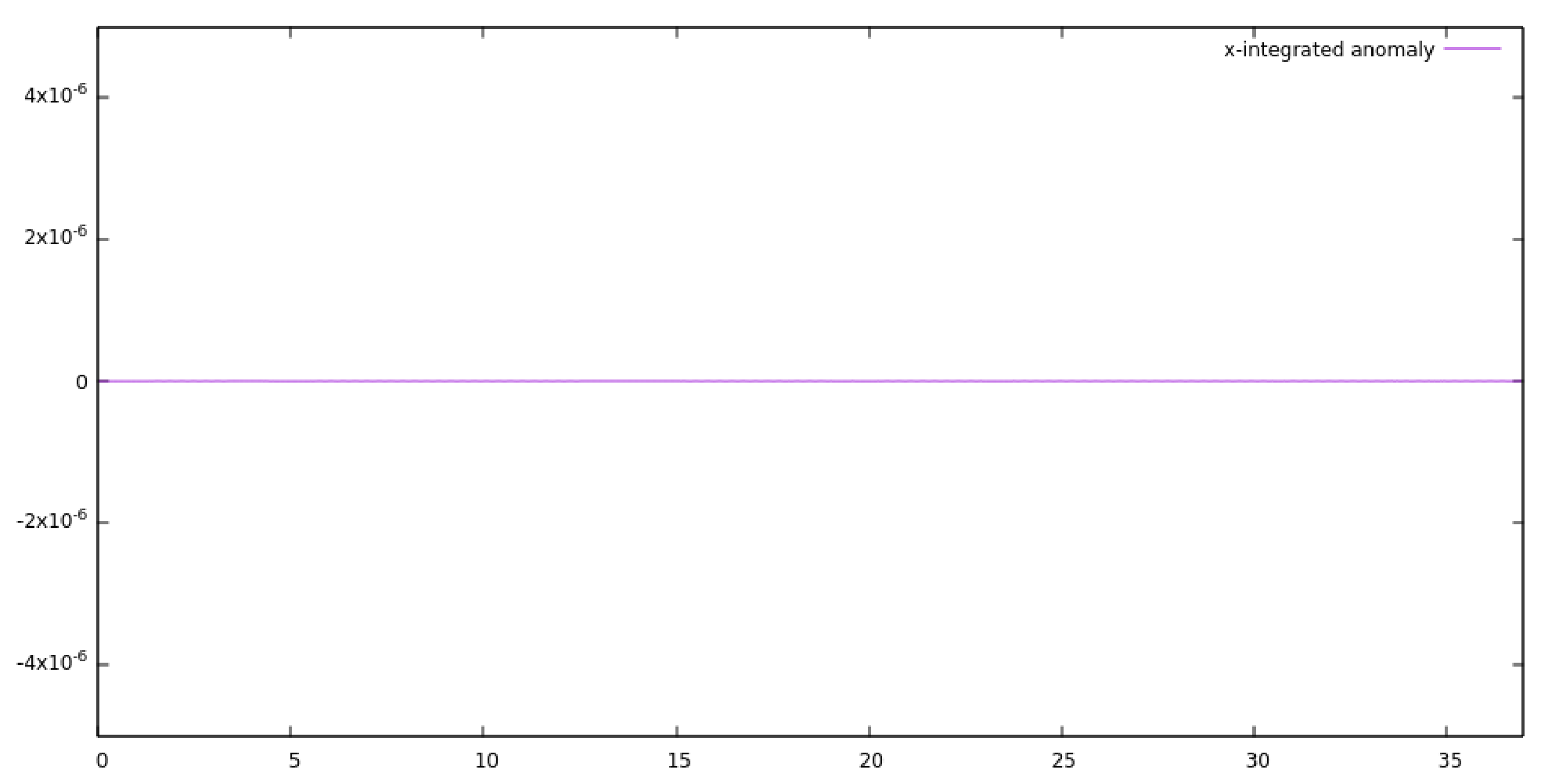}
\includegraphics[width=1cm,scale=2, angle=0,height=6cm]{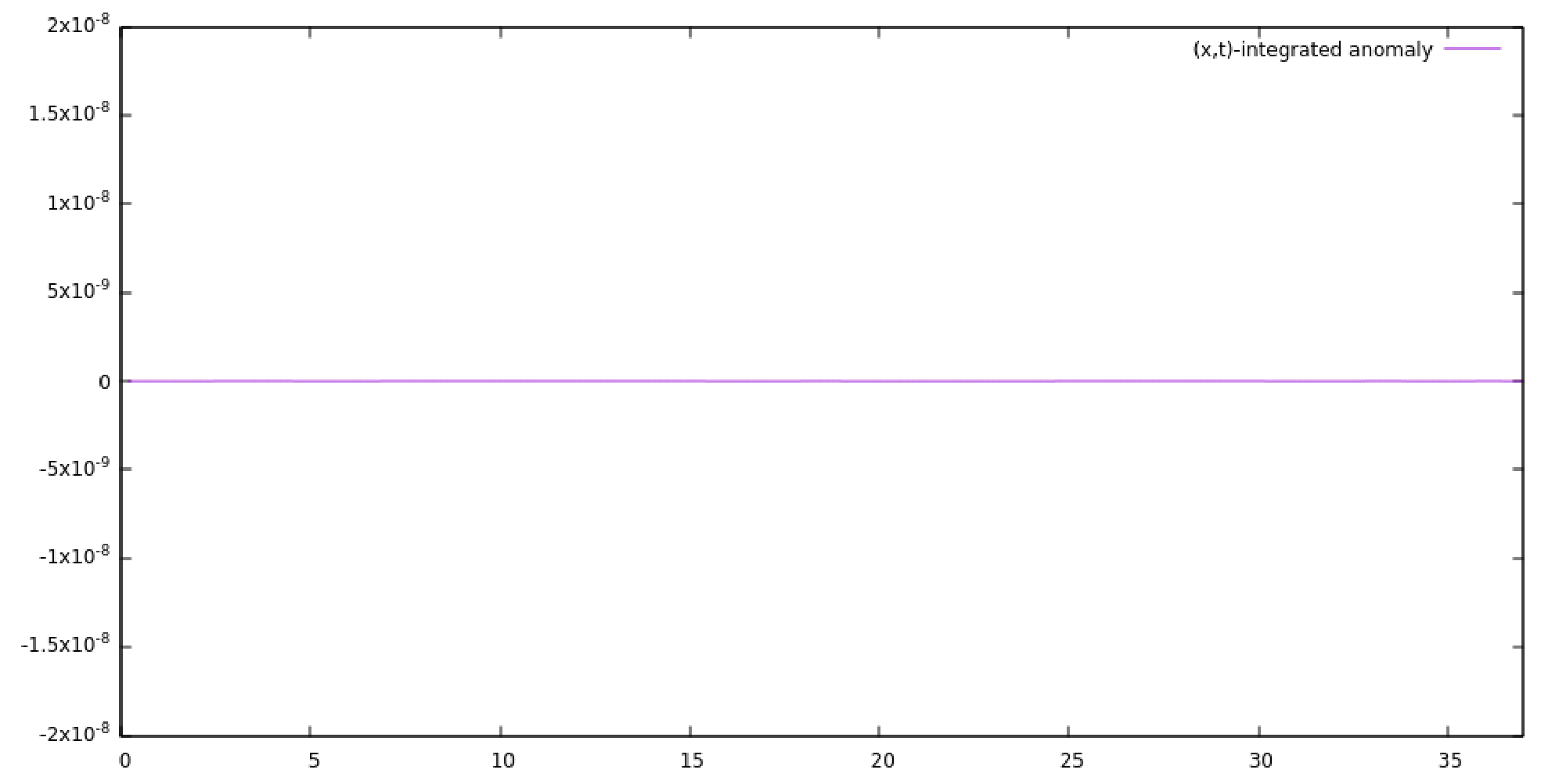}
\begin{center}
\parbox{5in}{\caption{(color online) Numerically simulated two-kink collision for $\epsilon_2= 0.9, \epsilon_1= 1.2$. The top Fig. shows the anomaly density $\hat{\a}$ in (\ref{nonlocano1}) for three successive times, initial (green), collision (orange) and final (magenta), respectively. The middle Fig. shows the $x-$integrated anomaly and the bottom shows the $(x,t)-$integrated anomaly. Notice that this figure vanishes within numerically accuracy $\approx 10^{-9}$.}}
\end{center}
\end{figure}

\section{Discussions and some conclusions}
\label{sec:discuss}

We have examined the quasi-integrability properties of a particular deformation of the potential KdV model (pKdV)  (\ref{dpkdv}) using a direct approach in order to construct the quasi-conservation laws, and through a numerical simulation of 2-kink collisions. For the undeformed pKdV ($\epsilon_1=\epsilon_2=0$) we have constructed the parity ${\cal P}$ symmetric N-soliton solutions, and shown that the $(x,t)-$integrated anomalies of the usual pKdV model vanish. As representative quasi-conservation laws of the pKdV model we have examined the quasi-conserved charges $q^{(5)}_1$ (\ref{charge51})  and  $q^{(5)}_3$ (\ref{charge52}) for $X=0$, and shown the vanishing of the common anomaly $-\a v_x^2 v_{4x}$ using its odd parity property under ${\cal P}$ symmetry.  In addition, we have discussed an anomaly cancellation mechanism in order to get the exact conserved charge  $q^{(5), \g}_{X=0}=q^{(5)}_1 - \g \, q^{(5)}_3$ in (\ref{q5gX0}) for the parameter value $\g=1$.

Some charges, such as  ${\cal Q}_1$ in (\ref{calQ11}), when evaluated for the composite field $q \sim `kink' + q_{clb}$ must be renormalized by subtructing the infinite contribution of the continuous linear background $q_{clb}$ in  (\ref{clb}), i.e. the kink charge ${\cal Q}_{1\,kink}$ must be renormalized. This is analog to the renormalization procedure in the defocusing NLS model in which the normalization and momentum  charges associated to the dark soliton itself have been renormalized by subtructing a continuous wave background \cite{jhep4}.

Taking into account that the 2-soliton (\ref{usolxt}) satisfies the parity symmetry (\ref{P2sol}) it has been shown the analytical vanishing of the anomalies. Therefore, it provides an analytical proof of the quasi-integrability of the pmRLW theory (\ref{dpkdve1e2}), i.e. a sub-model of deformed pKdV in the particular case $\epsilon_1=\epsilon_2=1$. So, this adds a new strong result on the analytical proof, not only numerical, of the quasi-integrability of a (non-integrable) theory, discussed in \cite{npb, jhep6} for deformed KdV models. 

We have considered numerically the behavior of the quasi-conservation laws (\ref{quasi133}) and  (\ref{nonloc1})  with the relevant anomaly densities $\a^{(5)}_4(x,t)$ in (\ref{ancharge53}) and $\hat{\a}(x,t)$ in (\ref{nonlocano1}), respectively. The Fig.1 (top right) and Fig. 2 (top) show the anomaly densities for three successive times, initial (green), collision (orange) and final (magenta), respectively. The figures show the $x-$integrated and the $(x,t)-$integrated anomalies. The $(x,t)-$integrated anomalies in the both cases vanish within numerical accuracy. So, one can regard the charges $q^{(5)}_{4, X}$ in (\ref{charge53}) and $\hat{q}$ in (\ref{nonlocharge1}) as asymptotically conserved charges within numerical accuracy, for the 2-kink collision.
 
Some qualitative aspects of the Liouville's theorem remain true for continuous integrable systems admitting a Lax pair representation \cite{das}. So, one must have an infinite number of conservation laws whose conserved charges are in involution \cite{das, faddeev}. In this context, the presence of the novel towers of asymptotically conserved charges as above, even in the standard pKdV model, are restricted to special field configurations satisfying the symmetry (\ref{parity1}) and (\ref{paritys1}). Therefore, one can not use these charges, even though they are infinitely many, in order to match to the number of degrees of freedom of the pKdV model. Of course, the true conserved charges hold for general field configurations, i.e. being solitonic or not. We believe, that an anomaly cancellation mechanism is required in order to get  the set of true conserved charges of the standard pKdV model. For example, in order to get an exact conservation law in (\ref{charge5di}) for $X=0$ one must have $\g = 1$, which is the condition for the anomaly cancellation. In fact, subtructing the quasi-conservation laws (\ref{quasi12}) and (\ref{quasi13}) (for $X=0$) one can get (\ref{charge5di}) to be an exact conservation law provided that $X=0$ and $\g=1$, since their anomalies $ -\a v_x^2 v_{4x}$ cancel to each other for $\g =1$. Of course, the  charge $q^{(5) \g=1}_{X=0}$ in (\ref{charge5di}) will be conserved for general solutions of the pKdV.

\end{document}